\documentclass[11pt]{article} 
\usepackage{color}
\usepackage{amssymb}

\title{Logical Aspects of Quantum (Non-)Individuality}
\author{{\sc D\' ecio Krause}\thanks{Leave from the Department of Philosophy, Federal University of Santa Catarina. Florian\' opolis, SC - Brazil, partially supported by CNPq.} \\{\small Wolfson College} \\ {\small University of Oxford} \\ 
{\small  {\tt deciokrause@gmail.com}}}

\newcommand{\igual}{=_{\small\mathsf{def}}}
\renewenvironment{enumerate}{\begin{list}{}{\rm \labelwidth 0mm
\leftmargin 5mm}} {\end{list}}

\begin{document}
\maketitle


\begin{abstract}
In this paper I consider some logical and mathematical aspects of 
the discussion of the identity and individuality of quantum entities. 
I shall point out that for some aspects of the discussion, the logical 
basis cannot be put aside; on the contrary, it leads us to unavoidable
conclusions which may have consequences in how we articulate cer- 
tain concepts related to quantum theory. Behind the discussion, there 
is a general argument which suggests the possibility of a metaphysics 
of non-individuals, based on a reasonable interpretation of quantum 
basic entities. I close the paper with a suggestion that consists in em- 
phasizing that quanta should be referred to by the cardinalities of the 
collections to which they belong, for which an adequate mathematical
framework seems to be possible.
\end{abstract}

\centerline{--------------------------------}
{\sf {\footnotesize ``The subject-predicate logic to which we are accustomed depends for its convenience upon the fact that  at the usual temperature of the earth there are approximately permanent `things'. This would not be true at the temperature of the sun, and is only roughly true at the temperature to which we are accustomed.''  (Russell [1957])}} 


\section{Introduction}
The philosophical discussion of aspects of quantum physics is one of the most fruitful sources of philosophical puzzles, without even involving us in sophisticated physics, or mathematics, or logic. Among them, there is of course the problem of indiscernible `particles', a question people such as van Fraassen have referred to as one of the three most important questions in this field (van Fraassen [1991], p.193). But in the general discussion, there are certain details usually not covered sufficiently, but which may matter (and I think that they really \emph{do} matter) for philosophical issues, as for instance those entailed by the logic and the mathematics used to formulate and to discuss the problem of individuation. As I shall try to make clear below, when we pay attention to such `logical details', other clusters of interesting philosophical puzzles appear, and  I emphasize some of them, although in a non-rigorous way. Really, I believe with Russell that, for certain kinds of discussions, we cannot do without logical precision,\footnote{In the same paper quoted above [1957], he said that ``I ($\ldots$) am persuaded that common speech  is full of vagueness and inaccuracy ($\ldots$) For technical purposes [philosophy of course included], technical languages differing from those of daily life are indispensable."} and the problem of individuation seems to me to be one of the most characteristic in this respect. 

We should acknowledge that while most papers consider logic only informally, some others mention  the underlying logical and mathematical framework where the discussion is performed but, in almost of these papers (really, in every one I know), what we find is a reference to first-order languages.  Sometimes the discussion of identity is taken as adequate if we assume merely what has been called `the Hilbert-Bernays approach', considered also by Quine ([1986], pp. 62-63), and more recently by Saunders (Saunders [2006], Muller \& Saunders [2008]).  Furthermore, when the necessary mathematics is to be mentioned, the references seem to be directed to first-order ZF (Zermelo-Fraenkel set theory with the axiom of choice, although this axiom is almost never mentioned). 

What is the importance of the underlying logic and mathematics to the consideration of these mentioned philosophical issues? My aim in this paper is to contribute to the discussion, or at least to make philosophers reflect on the subject, by emphasizing just the logical basis that underlies all the discussions of the identity and individuality of quantum objects at least.\footnote{Some points emphasized here are also presented in Ketland [2006], a paper I knew  only when the bulk of this paper had already been finished (I knew Saunders' and Muller \& Saunders' papers firstly as preprints). Anyway, the arguments I present here differ from Ketland's, for he focuses on structuralism. I think the two papers complement each other, each exploring details not covered by the other. Furthermore, let me say that I am using the word `object'  without any commitment to saying that quantum entities are `objects', for I ought to confess I don't know either  what an object is or,  worse, what a quantum object is. This is just a neutral term, used here in the same informal way  as `entity'; \textit{quantum} would perhaps be more appropriate.}

What I will emphasize is that the standard treatment given to identity by philosophers in general (since physicists seem not to be occupied with these matters) is only identity relative to a bundle of properties and/or relations, or indiscernibility (indistinguishability) with respect to some the properties and/or relations. These may be `intrinsic' properties, `physical properties', etc., all of them according to me representing a certain particular selection, whose reasons do not interest us here. Since I am not adopting here a extreme formalist position, my aim being to discuss physics,\footnote{But I think that perhaps there is no way of getting away from  this formal approach. As we are trying to develop in detail in a research program with Newton da Costa, the meaning even of the logical symbols is to be given by their formal linguistic aspects, more or less in Curry's sense [Curry 1966]. According to da Costa---informal conversation---, it is a myth to say that, for instance, Tarski's semantics fixes the meaning of the logical symbols (say, the logical connectives), for in the metamathematics we use the same symbols we are trying to give precision to.} in order to consider  the `true' identity relation (if there is one) we shall take into account also the metalanguage we use to consider semantics and meaning. Thus, I shall begin the next section by presenting some facts already well known in logic, but which will here be repeated in order to keep the paper suitable for the general reader, and in order also to shed some light on some aspects of the  metamathematics involved. After this, I shall show how the concept of identity is treated within classical logic and mathematics (what I call the `traditional' or `classical theory of identity'), in order to bring the light some (usually) `hidden' aspects of this theory. In this approach, the Principle of the Identity of Indiscernibles (PII) of course enters the scenario, and we shall consider some up-to-date discussions of its status. Finally, after my sketching my own point of view, I turn to Muller \& Saunders `relational'  entities to see what their definition entails. I shall maintain that relationals (in their sense) are also absolutely discernible (in their sense), and that, if we we keep the discussion within classical logic and mathematics, any entity will obey PII.

Let us begin by considering indiscernibility. 

\section{Indiscernibility in a structure}
Usually, people say that something (say, my pen) is an individual, for it can be distinguished from others even quite similar, perhaps by a scratch it has in distinction to any other. As we know, this is Leibniz's PII. But, as emphasized in French \& Krause [2006], we would not confuse distinguishability and individuality, for what distinguishes a thing from others may be not what makes it an individual. As  French \& Krause put it, there is 

\begin{quote}
$\ldots$ a conceptual distinction between \textit{individuality} and \textit{distinguishability} ($\ldots$) [w]e can approach the distinction by focusing on the particularity, of individuals, as contrasted with the universality, or non-instanciability, of their attributes. The former pertains to something, some `principle of individuality', as it is often called, which is `internal' or `particular' to the individual in the sense of being associated with it alone. The latter involves the individual's relationship with other particulars and is therefore `external' in some sense. One way of articulating this distinction is to imagine a possible world in which there is only one entity; this entity cannot be regarded as distinguishable from others---or so it is claimed---since there are no other entities, yet it may surely be considered an individual. (op.cit., p.6)
\end{quote}

Thus, they continue, we could take `individuality' to be that in virtue of which an individual is an individual, given, as usual, by some `principle of individuation', while by `distinguishability' we mean that in virtue of which an individual is distinguished from others. The search for a principle of individuation, as is well known (and not forgetten by these authors), pervades practically all the history of western philosophy. But present day physics seems to show us a different kind of entity, which can be taken as `more than one', apparently without any underlying principle of individuation. In other words, quantum mechanics suggests that there may be absolutely indistinguishable quanta, which cannot be distinguished by any means provided by the theory, and that even so cannot be reduced to one and the same entity; really, although indiscernible, they \textit{count} as more than one. As far as we know, there would be no quantum theory without indiscernibility.  

It is well known from the standard formalism how quantum mechanics treats this concept.  In this framework, indiscernibility is treated by assuming certain symmetry conditions (symmetric and anti-symmetric functions/vectors) that describe the relevant physical phenomena. Roughly speaking, in considering $n$ indiscernible quantum objects, we start by labeling them (that is, by considering them as individuals, identified by their labels), say by  $x_1, \ldots, x_n$, and then we use permutational symmetry to express the indiscernibility: if $F$ is an $n$-ary predicate, then $F(x_1, \ldots, x_n)$ is equivalent to $F(x_{\pi(1)}, \ldots, x_{\pi(n))}$ for every permutation $\pi$ defined on $\{1, \ldots, n\}$. That is, the invariance of permutations would be the tantamount of the indiscernibility of quantum entities. I intend to emphasize that this approach reflects indiscernibility as a relative concept only. To see that, and before going back to quantum issues, let us consider how mathematicians deal the concept of indiscernibility.
 
To exemplify, take the additive group of the integers, $\mathcal{Z} = \langle \mathbb{Z}, + \rangle$. This structure has two automorphisms (functions that `preserve' all the relations and operations of the structure), namely, the identity function $g(x)=x$ and the `opposite mapping'  $h(x)=-x$.\footnote{The favorite example in many papers are $i$ and $-i$ in the field of complex numbers, and the  automorphism,  apart from the identity function, is the function which associates with a complex number its conjugate.The arguments in the two cases are of course equivalent.} The collection of the automorphisms of a structure, taken with the binary operation of composition of functions, forms a group, the Galois group of the structure. Mathematicians say that two objects $a$ and $b$ in the domain of a certain structure $\mathcal{A}$ are indiscernible in $\mathcal{A}$ (or are $\mathcal{A}$-indiscernible) if there exists an automorphism $f$ of $\mathcal{A}$ such that $f(a)=b$. Thus, $2$ and $-2$ are $\mathcal{Z}$-indiscernible, although they are of course not identical. The relevant fact is that the difference between 2 and $-2$ cannot be seen \emph{from the inside} of the structure $\mathcal{Z}$.  Nevertheless this can be done  in a larger structure, say $\mathcal{Z}' = \langle \mathbb{Z}, +, \{n\}_{n \in \mathbb{Z}}\rangle$, obtained by adding to the original one all the singletons of the elements of $\mathbb{Z}$ (there are other ways to do it, but this one is enough for our purposes). In this extended structure, we can associate to each singleton $\{n\}$ the property ``to be identical to $n$'', and then 2 and $-2$ can be distinguished; for 2 is identical with 2 ($2 \in \{2\}$), but $-2$ is not ($-2 \notin \{2\}$). When the only automorphism of a structure is the identity function, as in  $\mathcal{Z}' = \langle \mathbb{Z}, +, \{n\}_{n \in \mathbb{Z}}\rangle$, we say that the structure is \textit{rigid}, that is, its Galois group is the trivial group. In our example, the structure $\mathcal{Z}$ was `extended' to a rigid one, namely $\mathcal{Z}'$ (we can in fact prove that $\mathcal{Z}'$ is rigid). 

What can we say of the general situation, that is, of any structure $\mathcal{A}$ whatever? Our first claim is: the answer depends on the background mathematics we are using. Without loss of generality, we can assume that all the mathematics we need can be performed in a set theory such as the Zermelo-Fraenkel system (perhaps with the axiom of choice, we shall assume here), which we shall consider as including the axiom of foundation---by the way, this is the theory that is undertood by most philosophers when they speak about ZF. In ZF, we can prove that \textit{every structure can be extended to a rigid structure}. This is an important result for it says that within the scope of standard mathematics (read: mathematics that can be `constructed' within ZF) we can always distinguish between two distinct objects whatever, if not in the  structure where they are being described, then in some of its rigid extensions. In other words, ZF, or perhaps we should say the well-founded universe $\mathcal{V} = \langle V, \in \rangle$, is rigid. Intuitively speaking, within ZF (and within the mathematics built in ZF), there are no \textit{genuine} indiscernible objects (objects which differ \textit{solo numero}), but  only objects indiscernible with respect to a structure. This is a `logical' fact we cannot avoid once we have assumed such a `classical framework'. 

\section{Enters quantum mechanics}
How does quantum theory enter into this kind of discussion? If we pay attention to the mathematical structure of quantum mechanics (QM for short,  and here we shall be concerned with non-relativistic quantum theory only),\footnote{The philosophical discussion of quantum field theories is already impressive, and we guess that what we are saying here (with due qualifications) can be extended even to relativistic quantum mechanics, for it too can be described within ZF.}  we can consider it as characterized by structures like 

\begin{equation}
\mathcal{Q} = \langle F, S, Q_0,\ldots,Q_n, \rho \rangle 
\end{equation}

\noindent where $F$ is a mathematical model of standard functional analysis, $S$ is a set of ``physical situations", $Q_0,\ldots, Q_n$ representing the observational part of the theory (physical observables) and $\rho$ is a mapping which ascribes to each $s \in S$ an adequate Hilbert space in $F$, and an Hermitian operator on that Hilbert space to each $Q_i$, $i = 1, \ldots, n$  (Dalla Chiara and Toraldo di Francia [1981], pp. 85ff). Certain postulates, well known from the corresponding literature, an be taken as understood. Structures of this kind can be axiomatized within ZF by a set-theoretical predicate in the sense of Suppes (Suppes [2002]), and it copes with the main mathematical traits of non-relativistic quantum theory, encompassing the standard formalism of QM \textit{via} Hilbert spaces.

But it is important to remark that the postulates of QM are not only these `specific' of `factual' postulates, which have (we suppose) physical significance, but  also those of ZF and, hence, also those of first-order logic (we are supposing ZF as a first-order theory, yet we are not sure that this is the best choice, as we shall remark below).\footnote{Perhaps a quantum field theory, encompassing particle physics and gravitation, cannot be easily described this way, for it seems to have no way of making the base theories compatible, without resorting to a non-classical logic that tolerates contradictions---that is, a paraconsistent logic. But this is a guess. See Krause \& Bueno [2007].} The same of course happens with, say, groups in standard mathematics: beyond the group postulates (associativity, identity element and the existence of inverses), we ought to acknowledge, there are the ZF postulates (which give us concepts such as binary operation, cartesian product, etc.) and those of logic, which enable us to make deductions and prove the relevant theorems. Thus, we can perhaps speak here of a kind of Duhem--Quine thesis: if some result contradicts the specific axioms, it contradicts the whole system, hence the logic and mathematical axioms as well. More directly stated, no consequences of quantum mechanics can contradict its underlying logic and mathematics unless the specific axioms are inconsistent with the (supposed) consistent logic and mathematical axioms. Thus, as we shall see below, there can be no `legitimate' indistinguishable objects. Let us continue.  

$\mathcal{Q}$ is a structure in ZF, so, although it might be hard to prove, if its Galois group is not trivial (I don't know if there is already an explicit  proof of this fact), then it can be extended to a rigid structure where all (the representable) objects are individuals, in the sense of obeying the classical theory of identity embedded in ZF. In short, we can say in a fragment of ZF used to cope with QM that some entities are indiscernible, physically indiscernible if you wish, or something like that, but owing to the commitment of the theory to classical logic, there is no escape: all entities are individuals in the sense explained above.\footnote{Perhaps this distinction may be used to give an interpretation to van Fraassen's way of `saving'  PII. In his [1991], p.376, he  notes: ``identical particles ($\ldots$) are certainly qualitatively the same, in all the respects represented in quantum-mechanical models---yet still numerically distinct''. In a previous paper, he was still more explicit, insisting that ``if  two particles are of the same kind, and have the same state of motion, nothing in the quantum-mechanical description distinguishes them. Yet this is possible" [1984]. In Krause \& Bueno [2007], it was suggested, following the parallel case of `Skolem's paradox' in set theory, that the particles might be distinguished \textit{outside} the framework of quantum mechanics (if this is so, would it be an `incomplete' theory?). The same perhaps can be said in the general case we are discussing here.}
 Thus, we are faced with a dilemma (which of course does not bore everyone, but surely would bore physicists): to continue using standard classical languages at the expenses of the necessity of using symmetry conditions, as usual, or to go deep into the philosophical and foundational problem of finding an adequate language to express the fact that we really should begin with indiscernible entities in the first place.\footnote{This is precisely the proposal of Heinz Post and it is implicit in what it was termed the Manin Problem in French \& Krause op.cit., where further references are given. A  first attempt at constructing an adequate language can be seen in Domenech et. al [2008].} 
 
We note in brief that this discussion may be relevant also for  QFT (although this topic would deserve further study). For instance, we find Stachel saying that ``from the perspective of relativistic field theory ($\ldots$) one cannot attribute individuality to units that are truly field quanta ($\ldots$) [which] manifest no inherent individuality" (Stachel [2005]). So, as far as in QFT we are working within classical logic, it seems that there is no scape: although we acknowledge that 'particles' are epiphenomena of fields (Falkenburg [2007]), they are also individuals. 

These remarks (I hope) show that, strictly speaking,  when we really pay attention to the logical and mathematical framework we are working within when \textit{doing} theoretical physics, interesting and important questions arise when we think of the standard philosophical discourse. Let me quote a few of them:\footnote{Newton da Costa---private conversation---reminded me of some  interesting questions on this respect. Some of them I am mentioning here together with my own questionings, but I am  putting them in my own words.} is there a model of (first-order) ZF that copes with reality? If so, there is a countable model too, indeed models of every infinite cardinality since  the L\"owenheim--Skolem theorem holds for \textit{any} first-order theory.\footnote{We recall that, for first-order logic with equality, the theorem says only that if there is a model then there is a model that its \textit{at most} denumerable. It is not impossible that all models are finite. ZF, of course, has no finite models.} Furthermore, set theories such as ZF don't have intuitive models, unless we are strong platonists like G\"odel. So, if  ZF has no intuitive model, how can such a theory provide us an intuitive model of reality? Or, since we cannot characterize a standard model of (first-order) ZF (one where identity is `true' identity), how can we be aware that the natural numbers, which we use, for instance, in the Fock space formalism, are really finite? (The notion of cardinal is not \textit{absolute}---in Skolem's sense---but depends on the model.) More: recall that in the standard Hilbert space formalism, we deal with bases for the relevant Hilbert spaces. More specifically, we deal with orthonormal bases formed by eigenvectors of certain Hermitian operators. This is possible because we can prove, using the axiom of choice (which is part of the background theory used here, at least by hypothesis) that any Hilbert space \emph{H} has a basis. Moreover, it can also be shown that each basis has a specific cardinality, which is the same for all bases of \emph{H} (this is defined to be the dimension of the space). But in certain models of set theories in which the axiom of choice does not hold in full generality (but there is some other axiom instead---although we shall not mention the details here), such as in La\"uchli's permutation models, we can obtain: (a) vector spaces with no basis, and (b) vector spaces that have two bases of different cardinalities (Jech [1991], p. 366). Now, if a vector space has no basis, it is supposed that it cannot be used as part of the standard formalism of quantum mechanics, for the latter formalism presupposes the availability of suitable bases. As a result, we see that the formalism depends crucially on the background mathematics.

Let us take another example (also taken from Krause \& Bueno [2007]): we know that an important concept in quantum mechanics is that of an unbounded operator, such as the position and momentum operators.\footnote{If $\hat{A}$ is an operator, then $\hat{A}$ is bounded if for any $M > 0$ there exists  a vector $\alpha$ such that $\| \hat{A}(\alpha)\| \geqslant M \|\alpha\|$.} However, consider the theory ZF+DC, where DC stands for a weakened form of the axiom of choice  entailing that a `denumerable' form of the axiom of choice can be obtained.\footnote{In particular, if $\{B_n : n \in \omega \}$ is a countable collection of nonempty sets, then it follows from DC that there exists a choice function $f$ with domain $\omega$ such that $f(n) \in B_n$ for each $n \in \omega$.} Then it can be proven, as Solovay showed, that in ZF+DC (which is supposed to be consistent) the proposition ``Every subset of $\mathbb{R}$ is Lebesgue measurable" cannot be disproved. This proposition is false in standard ZF with the axiom of choice. Furthermore, in Solovay's model, we can prove that each linear operator on a Hilbert space is bounded. Thus, taking into account these metamathematical results, can we use  Solovay's model in ZF+DC to elaborate our favorite quantum mechanics? Since we need unbounded operators, we think that perhaps we cannot. So, the choice of a suitable background mathematics is crucial here too.

What are we to say about individuality?  In order to consider it, we shall begin by casting an eye on the classical theory of identity.

\section{The classical theory of identity}
Intuitively speaking, identity is a relation that an object has with itself and with nothing else. Of course this `definition' is circular, and perhaps a watertight definition of identity cannot be achieved. So, we shall follow the standard procedure and say that $a=b$ means that the objects denoted by $a$ and by $b$ are \textit{the very same} object, whatever that means. Set-theoretically (in extensional set theories), the identity relation on (the set) $A$ is the binary relation (called the \textit{diagonal} of $A$):

\begin{equation}
\Delta_A \igual \{\langle x, x \rangle : x \in A\}. 
\end{equation} 

Thus we can transform the difficulties in defining identity into another question, which sounds quite natural within this mathematical context: can we axiomatize $\Delta_A$? The answer is that it depends on the language employed. Let us consider in turn the best known possibilities available.

\subsection{First-order languages} 
The usual treatment of identity in first-order languages is to consider it as a primitive concept. So, being $=$ a primitive binary predicate symbol, we can use the so-called Frege axioms: 

\begin{enumerate}
\item ($=_1$) \, $\forall x (x=x)$
\item ($=_2$) \, $\forall x \forall y (x = y \to (\alpha(x) \to \alpha(y))$, where $\alpha(x)$ is any formula where $x$ appears free and $\alpha(y)$ is obtained from $\alpha(x)$ by the substitution of some  (not necessarily all) free occurrences of $x$ by $y$, and $x$ is free for $y$ in $\alpha(x)$. 
\end{enumerate}

The obvious purpose of these postulates (as we understand them semantically today) is to fix the idea that when we prove $a=b$, we are saying that $a$ and $b$ are the very same object (or referent). Thus, there appears a link with semantics, and we need to consider it.\footnote{By the way, as far as the expression `classical logic'  is not completely clear, if we accept a logical system formulated in a `classical'  way, say formal arithmetic, but do not accept standard semantics, done within a standard set theory, can we still speak of classical logic? I shall assume that classical logic encompasses `classical' semantics. Full stop.} We must acknowledge that, when our language is interpreted in a structure $\mathfrak{A} = \langle D, \rho\rangle$ with domain $D$ ($D \not= \emptyset$), these axioms do not axiomatize identity, in the sense of not fixing the diagonal of the domain $D$. They axiomatize a congruence relation, but not necessarily identity. Why? 

Suppose that $\mathfrak{A} = \langle D, \rho\rangle$ is, as above, an interpretation for our first-order language (encompassing identity), where $D\not= 0$ and $\rho$ is the denotation function. Let us call $\Delta_D$ the diagonal of $D$. Now let $\sim$ be any congruence relation defined on $D$, and let us take the set $D^* \igual D/\!\sim$ as the domain of a new interpretation $\mathfrak{A}^* = \langle D^*, \rho^*\rangle$  for $\mathcal{L}$. We can prove  that there exists a mapping $f : D \mapsto D^*$ defined as follows (I shall assume for simplicity that $\mathcal{L}$ has no functional symbols):

\begin{enumerate}
\item (i) For every $x \in D$, $f(x)$ is the equivalence class in $D^*$ to which $x$ belongs.
\item (ii) For every $x$ and $y$ in $D$, $\langle f(x), f(y) \rangle \in \Delta_{D^*}$ iff $\langle x, y \rangle \in \Delta_D$. 
\item (iii)  For every $n$-ary predicate symbol $P$ of $\mathcal{L}$, $\langle f(x_1),\ldots,f(x_n)\rangle \in \rho(P)$ iff $\langle f(x_1),\ldots,f(x_n)\rangle \in \rho^*(P)$
\item (iv) For every individual constant $a$, $\rho^*(a) = f(\rho(a))$.
\end{enumerate}

Then, we can prove (Hodges [1983], p. 64) that the structures $\mathfrak{A}$ and
$\mathfrak{A}^*$ are \textit{elementary
equivalent},  that is, intuitively speaking, the sentences of $\mathcal{L}$  which are true in $\mathfrak{A}$ are the same as those which
are true in $\mathfrak{A}^*$. In other words,
the language  $\mathcal{L}$ confuses us whether we are speaking of individuals (elements of $D$) or classes of individuals (elements of $D^*$).

Let us think a little bit about this result. It says that whatever sentence
$\alpha(y_{1}, \ldots, y_{n})$ of $\mathcal{L}$  is such that

\begin{equation}\label{true}
\mathfrak{A} \models \alpha(y_{1}/x_{1}, \ldots, y_{n}/x_{n}) \, \,
 {\rm iff}   \, \,  \mathfrak{A}^* \models \alpha(y_{1}/f(x_{1}), \ldots,
y_{n}/f(x_{n})).
\end{equation}

Thus, $\alpha(y_{1}, \ldots, y_{n})$ is true according to
one interpretation if and only if it is true according to the other.

So, from the point of view of the first order language $\mathcal{L}$, it is not
possible to know whether we are dealing with an element of $D$ or
with an equivalence class of $D^*$ (hence, a certain
collection of elements of $D$). 

Thus, we ought to conclude that first-order languages do not distinguish individuals from collections of individuals (say an electron---taken as such---from a collection of electrons). It is of some interest that the same kind of conclusion is reached when we use group theory to classify elementary particles: we arrive at most at \emph{kinds} of particles (Castellani [1998]). 

\subsection{Defining identity}
In higher-order languages (say, second-order, which suffices for the argument) we can define identity by means of Leibniz's Law:\footnote{Some authors prefer to call Leibniz's Law what we are calling here the Indiscernibility of Identicals (II). We reserve the term `Leibniz's Law'  for (\ref{LL}).} 

\begin{equation}\label{LL}
x=y \igual \forall F (F(x) \leftrightarrow F(y)),
\end{equation}

\noindent where $x$ and $y$ are individual variables and $F$ is a variable ranging the set of the properties of the individuals.\footnote{Including those which $x$ and $y$ do not possess.} This definition is based on Whitehead and Russell's \textit{Principia Mathematica}, and encompasses both PII (see below, (\ref{PII})), and its converse, the Principle of Indiscernibility of Identicals (II).  If we admit in the range of the quantifier  the property ``to be identical with $a$", that is, $I_a(x) \leftrightarrow x=a$, we don't need the bi-conditional and PII below becomes a theorem of higher order logic, that is,

\begin{equation}\label{PII}
 \forall F (F(x) \rightarrow F(y)) \to x=y.
\end{equation}

The proof can be given as follows, just to insist on the main point (see French \& Krause [2006], p.255).  Let us consider an interpretation ${\mathfrak A}$ such
that ${\mathfrak A} \models a = b$, that is, $\langle \rho
(a), \rho (b) \rangle \in \Delta_{D}$, where $D$ is the domain of
${\mathfrak A}$. Then, for every unary
predicate $F$, ${\mathfrak A} \models F(a) \to F(b)$, hence, by
Generalization,
${\mathfrak A} \models \forall X (X(a) \to X(b))$. Conversely, let us
assume that ${\mathfrak A} \models \forall X (X(a) \to X(b))$. Hence,
since $I_{a}(x) =_{\mathsf {df}} x = a$, then ${\mathfrak A}
\models I_{a}(a) \to I_{a}(b)$. But, since ${\mathcal A} \models
I_{a}(a)$, as every object is identical to itself, so also
${\mathfrak A} \models I_{a}(b)$. But then $\rho (a) = \rho (b)$ and
therefore ${\mathfrak A} \models a = b$.  This is another interesting thing about the philosophical discussion of the subject. Usually, PII is formulated using material implication, but in doing that, philosophers are implicitly assuming the predicates $I_a$, that is, classical identity theory. 

Higher-order languages are of course stronger than first-order languages. In particular, (\ref{LL}) defines identity  and not merely a congruence relation, as in first-order logic. But in order to ensure that Leibniz's Law grants that two individuals $a$ and $b$ are \textit{really} identical (refer to ``the same'' object), we need to work with full (standard) models, that is, structures that encompass \textit{all} the subsets of the domain of the interpretation. If not, that is, if we consider Henkin-style models, it is easy to show that we can present situations where $a$ and $b$ satisfy (\ref{LL})  and even so are distinct (French \& Krause op.cit., p.257). For instance, let us take a domain $D = \{1,2,3,4\}$ and suppose that the language has three unary predicates which are interpreted by the subsets $\{1,2\}$, $\{1,2,3\}$ and
 $\{1,2,4\}$. Furthermore, interpret $a$ as  $1$ and $b$ as $2$. Then, of course $a$ and $b$ belong to the same sets (of the structure), that is, satisfy the same predicates, thus obeying (\ref{LL}), but  $1$ is not \textit{identical} with $2$.

In the standard extensional set theories, such as first-order ZF (without \textit{Urelemente}), in addition to the postulates ($=_1$) and ($=_2$), we add the Axiom of Extensionality:

\begin{equation}
\forall x \forall y (\forall z (z \in x \leftrightarrow z \in y) \to x=y).
\end{equation}

Thus, sets are identical iff they have the same elements (`same' in the sense of the above Frege postulates ($=_1$) and ($=_2$) plus extensionality). If the theory encompasses \textit{Urelemente}, a slight qualification must be introduced, and the result is that  atoms will be identical when they belong to the same sets.\footnote{Important to remark that the axioms---that is, extensionality---are such that the theory enables the existence of atoms distinct from the empty set.} (Even if we think or permutation models in the sense of Fraenkel--Mostowski---Krivine [1973], where \textit{Urelemente} are indiscernible, this is again a kind of restriction to certain structures, namely, the permutation models, for we can form the singleton $\{a\}$ of any \textit{Urelemente} $a$, which, as we have shown, makes it an individual in the whole universe os sets.)

Can we say that the ZF postulates define identity? Yes and no. Yes, for they imply that sets with the same elements are identical, and we know what `the same elements' means: they belong to the same sets. But, recalling what was said above, first-order set theory (recall that ZF is our paradigm), if consistent, has an infinite model. Thus, by the L\"owenheim--Skolem theorem, it has also a denumerable model, and it has also models of any infinite cardinality, as we have mentioned, and we cannot know which model we are speaking about. Well, you can say: fix on a model. But in regard to physical applications, we can answer: which one? And, once we have chosen one model, we can ask: why `this' one and not any other? Of course philosophical discussion enters also here, perhaps by pointing out, the needs for pragmatic criteria in choosing a model. 

Back to the point, we realize that the above characterization of identity shows that every object in the well-founded von Neumann universe $\mathcal{V} = \langle V, \in \rangle$ an individual. Really, just recall the basic argument, for any set $a$, we can form the set $\{a\}$ (by the pair axiom), which can be identified as the extension of the predicate``to be identical to $a$".  Thus, the only object of the universe of sets that satisfies this predicate is $a$ itself, and any other object is \textit{distinct} from it. 

\subsection{First-order languages, again}
Although, as we have seen, the statement of PII needs at least a second-order logic, it is quite common to find philosophers discussing this principle even when they give the idea that they have restricted the discussion to first-order languages only.\footnote{For instance, Muller \& Saunders fix $\mathcal{L}_{\mathrm{QM}}$, `the language of QM', as a first-order language (p.520), but they then quantify over properties and relations (p.528).}
Let us then consider briefly PII within the context of  first-order logic.   A way to express PII in denumerable first-order languages  is to consider the converse of ($=_2$). I note that such a converse is not a theorem of first-order logic. Suppose we add it as an additional axiom: the most we can do is to write it as a denumerable schema, such as 

\begin{equation}\label{1PII}
\forall x \forall x ((\alpha(x) \to \alpha(y) \to x=y)
\end{equation}

What does it mean? Let us suppose a particular case, without losing the generality of the conclusion, taking atomic formulas with unary predicates only, where $F$ is an unary predicate symbol. Thus (\ref{1PII}) reads (without quantifying over $F$, for our language $\mathcal{L}$ is a first-order language): 

\begin{equation}\label{2PII}
\forall x \forall y (F(x) \to F(y)) \to x=y
\end{equation}

Semantically, suppose that our domain is the set of natural numbers $\omega$. Then, according to standard semantics, to each $F$ we associate a subset of $\omega$.  But the number of subsets covered by (\ref{2PII}) is at most $\aleph_0$, while $\omega$ has $2^{\aleph_0}$ subsets! Something is missed here  (some kind of hidden properties not made explicit in the language?), that is, $x$ and $y$ can satisfy all the $F$s encompassed by the schema  (\ref{2PII}) and even so not be the same natural numbers. 

This shows that if we use the above schema to represent PII, we cannot ensure that the objects that obey exactly the same predicates or formulas (which are at most denumerable in number) are really identical. And this remains true for whichever interpretation we intend to give to the objects: `relational objects', `physical indiscernible systems', etc. (see below). Any one of these definitions expresses indiscernibility only relative to a certain denumerable bundle of attributes. 

In order to provide real identification (mathematically speaking), we need to take \textit{all} the predicates in order to be able to consider \textit{all} subsets of the domain and then apply the set-theoretical postulates  to see whether the objects are really identical. Thus, conclusions like that of Muller \& Saunders, who assert that the indiscernibility of $a$ and $b$  by the predicate $=$ axiomatized by Frege's axioms is equivalent to the identity of $a$ and $b$ ($a=b$), need care, for it depends on what we understand by $a=b$. If this is  taken as meaning that $a$ and $b$ are the very same object, then the conclusion may be  false, as we have saen, since Frege's axioms do not characterize the diagonal of the domain. The same happens with respect to their concept of physically indiscernible objects, which they use as `PII in QM' (p.523 of their paper): if two physical systems are indiscernible with respect to a subset of predicates of the language of QM,\footnote{To make the point again, it is not clear to me what this language is:  can QM be formulated as a first-order theory as they seem to suggest? I doubt it. We need some strong set-theoretical resources, which go beyond first-order logic, although ever since Skolem we have known how to axiomatize set theory as a first-order theory. This is another point,  still not made clear in the literature, that deserves the consideration by philosophers of science.} then they are identical ($a=b$). This can make sense from the physical point of view, but as far as the entities are described within ZF, physical indistinguishability does not entail identity. Further, even when we say that a permutation of indiscernible quantum objects does not conduce to a different situation, we should add 
`physical situation', for from the perspective of classical logic of course changes were made by the simple fact that we may be exchanging two distinct objects, which violates the axiom of extensionality. I guess that it is really quite difficult to reconcile the language of quantum physics, once it supposes indiscernibility as something not made by hand, with that of classical logic.\footnote{Perhaps this was the reason Yuri Manin said that quantum mechanics has no language of its own, and has to make do with a fragment of standard functional analysis. See French \& Krause op.cit., pp.239ff.}

Of course this kind of discussion opens the door to other forms of interpretation, such as those which prefer to speak of processes, events, or something similar. But from the logical and mathematical point of view, until now and as far as I know, no clear characterization in logical terms has been achieved by these proposals. The same can be said concerning structural realism. There, instead of speaking of `objects' having properties or entering in relations, we should speak first of all of relations. Although I don't know a definition of structure which fits such claims, one that enables us to speak of relations without the relata, this alternative may be interesting (see Krause [2005]). But let us go back to our point. 

Even in first order languages, we can try to define identity by finding  a formula $\alpha(x,y)$ on two free variables $x$ and $y$ for which we can state 

\begin{equation}
x=y \igual \alpha(x,y).
\end{equation}

This is what is appropriate, some people maintain, when we have a finite number of predicates. For instance, if our language has only the binary predicate $P$ and the unary predicate $Q$, then we can take $\alpha(x,y)$ to be 

\begin{equation}\label{HB}
\forall z ((P(x,z) \leftrightarrow P(y,z)) \wedge (P(z,x) \leftrightarrow P(z,y))) \wedge (Q(x) \leftrightarrow Q(y)).
\end{equation}
 
This definition is attributed to Hilbert \& Bernays, and used by Quine [1986], chap.5, and Saunders [2006]. Does it define identity (in the sense of fixing the diagonal of the domain of an interpretation)? Not really. It defines indiscernibility with respect to the chosen (considered) predicates of the language. Quine himself acknowledges this: 

\begin{quote}
 [i]t may happen that the objects intended as values of the variables of quantification are not completely distinguishable from one another by the ($\ldots$) predicates. When this happens, (3) [his equation for (\ref{HB}) above] fails to define genuine identity. Still, such failure remains unobservable from within the language.  (Quine [1986], p.63)
\end{quote}

We can see from the above discussion that if we consider semantics, the defined identity relation does not in general coincide (extensionally) with the diagonal of the domain. There is indeed no way to get the diagonal, but  to quantify over all subsets of the domain; thus we need higher-order logic (or set theory, which does that, but differently). 

\section{Quantum philosophical issues}
We have seen that the concept of identity and the idea of quanta as individuals raise lots of logical and philosophical problems. Thus, we might ask:  when investigating  quantum philosophical issues, why should we be so reluctant to consider a different alternative, say that of allowing that quanta to be non-individuals (entities which do not obey the classical theory of identity)? 

Non-individuals would provide a really good metaphysical interpretation of quanta. We could retain realism if we wished, for we can be realistic about of a kind of entity, non-individuals, which we can admit exist antecedently  to the relations holding among them. But we could also be ontological structuralists of a kind.  In Krause [2005], an attempt was made to develop, on the basis of quasi-set theory (French \& Krause op.cit., chap.7), a definition of structures whose relations do not depend on the (specific) relata. The main idea is that for each relation $R$ (binary, say, and distinct from membership),  if $R(a,b)$  holds, than $R(a',b')$ also holds, where $a'$ is indistinguishable from $a$ and $b'$ is indistinguishable from $b$. Thus, the variables $a$ and $b$ act as placeholders for (indexicals of) entities of certain sorts, as in chemistry, where it is indifferent what particular hydrogen atoms form a molecule of water, it being important only that they are hydrogen atoms. Of course this alternative presupposes objects (non-individuals) that enter into the relations, but not `specific' entities, that is, individuals. However, this is achieved,  we are committed to entities that can be classified (even if only primitively) as being of different kinds, and are capable, in certain circumstances, of being absolutely indiscernible. With the adequate mathematics (quasi-set theory may be such a theory), they can be put into collections with definite cardinalities (although without an associated ordinal). Furthermore, with such  mathematical formalism at hand, which may enable us to form collections of them, these indiscernible objects can be values of variables of an adequately regimented language; in other terms, they may \textit{exist} in the Quinean sense. 

This possibility  leads us directly to a bottom-up approach: instead of starting with the usual intuitive conception of macroscopic objects, and classical logico-mathematical frameworks, and going downwards to the loss of individuality, we could start with indiscernible entities at the bottom and try to build a corresponding quantum theory, coping with the main traits of the standard one, and of course getting the same results. Perhaps some of the philosophical puzzles disappear. But this is still a guess, so I should be silent concerning the consequences. Anyway, as we said before, the first part, namely, the initial `construction' of a quantum theory with non-individuals as primitive entities was proposed in Domenech et al. [2008], [2008a].  

Let us close with another general discussion on the identity of indiscernibles. 

\section{General discussion: relational objects}
The philosophical literature has distinguished various forms of PII, depending on the domain of the universal quantifier in (\ref{PII}). For instance, we can distinguish among PII(1), where the range of the quantifier encompasses all properties and relations of the objects, PII(2) where spatio-temporal relations are not taken into account, and PII(3), which admits only monadic properties (see French \& Krause [2006], p.40). But here I shall follow Muller \& Saunders in speaking of PII-A, the Principle of Absolute Indiscernibles, PII-R, the Principle of Relational Indiscernibles and PII, the Principle of Identity of Indiscernibles, \textit{tout court}. PII-A says that no two distinct objects can be \textit{absolutely} indiscernible, that is, there is always some property (represented by a monadic predicate) that one of them has but the other does not (cf. Muller \& Saunders [2008], p.503). PII-R says that no two distinct objects are relational indiscernible, that is, there is always a relational property that distinguishes them, while PII says that no two objects are absolutely and relationally indiscernible (Muller \& Saunders present the rigorous definitions of course, but use quantification over predicates and relations). Then they argue that PII-A $\rightarrow$ PII, so as PII-R $\rightarrow$ PII and PII $\leftrightarrow$ PII-R. All of this is quite intuitive given the above definitions. But then they say that PII-A is not necessary for PII, and that we should have $\neg$(PII $\rightarrow$ PII-A), and so it may be the case that PII $\wedge$ $\neg$PII-A. I think that, given the underlying logic and mathematical apparatus, this is a misleading way to putting things. Let us see why and what may be the consequences.  Firstly, recall that that two objects are weakly discernible if they obey an irreflexive and symmetric relation. This enable us to introduce a Principle of Weak Indiscernibility (PII-W) stating that no two distinct objects are weakly indiscernible.  Thus, summarizing Muller \& Saunders' argument, we have: 


\begin{enumerate}
\item (i) PII-A $\rightarrow$ PII (definition)
\item (ii) PII-R $\rightarrow$ PII (idem)
\item (iii) PII-W $\to$ PII-R (idem)
\item (iv) PII-A $\rightarrow$ PII-W (for absolute discernible objects are always weakly discernible---Muller \& Saunders [2008], p.529) 
\item (v) PII-A $\to$ PII-R ((iii), (iv), and propositional logic)
\item (vi) PII $\leftrightarrow$ PII-A $\vee$ PII-R (definition)
\item (vii) Hence PII $\leftrightarrow$ PII-R (by (v), (vi), and propositional logic)
\item (viii) Muller \& Saunders claim that $\neg$(PII $\rightarrow$ PII-A)
\item (ix) If (viii) holds, then PII $\wedge$ $\neg$PII-A
\item (x) So, PII-R $\wedge$ $\neg$PII-A (by vii and ix).
\end{enumerate}

But it is easy to see that PII-R $\rightarrow$ PII-A, so the three principles, PII, PII-A, and PII-R, are equivalent. The first argument comes from the above result about the extension in ZF of any structure  to a rigid structure. Then, even the relational structure considered in order  to say that the objects are relational discernible, can be extended to a rigid one encompassing (perhaps all) monadic predicates as well. Of course there are, in general, various ways to extend a structure to a rigid one, but what is important is that there is \textit{that} way involving the monadic predicates. The only way to avoid this conclusion is to keep to the relational structure, say by insisting that only `genuine' relational relations are important, but this is precisely what we are emphasizing physicists usually do. Really, any restriction of PII means precisely that we are building walls around ourselves in order to be confined to a structure. But in ZF there are no perfect walls. We can always escape and see the truth. The second argument (really, the same one but written differently) runs as follows, and shows that given $a \not= b$, there is always a set to which $a$ belongs but to which $b$ does not belong.\footnote{I am presenting an argumentation that emphasizes the role played by the property of self-identity.} Really, if there exists $R$ such that $R(a,b)$, this means that $\langle a,b \rangle \in R$ (recall that we are, by hypothesis, working in an extensional set theory), that is, $\{\{a\},\{a,b\}\} \in R$. So $\{a\} \in \bigcup R$ (the union of $R$, which can always e formed within ZF). Since we can always get $\bigcup R$, we arrive at the \textit{property} ``being identical with $a$" (self-identity), which we can define as $I_a(x) \igual x=a$. Since this can be done for every individual of the (finite) domain, we get the monadic properties that make them individuals; hence, if  $a \not= b$, for sure $I_a(a)$ but $\neg I_a(b)$. So, PII-A holds. 

It is important to remark that one can always say that the language has some chosen predicates and/or relations, and it is relative only to them that the considered entities would be `identical'. But what the above argument shows is that for finite domains (which seems to be the case of any quantum theory) we can always extend such a language to a language that encompasses the identity conditions for the elements of the domain (their ``self-identity predicates''); that is, to each $a \in D$ ($D$ being the domain), we can add to the language the predicate $I_a$ defined above. It is clear that this kind of extension leads to a conservative extension of the previous theory.\footnote{We say that the theory $T_2$ is a conservative extension of $T_1$ if the language of $T_2$ has all the symbols of $T_1$ (and possibly others) and is such that (1) every theorem of $T_1$ is a theorem of $T_2$ and (2) those theorems of $T_2$ that are in the language ot $T_1$ are also theorems of $T_1$.} We remark that this is always possible for finite domains which are \emph{sets} (collections of individuals). 

What about QM? You could say that, by using the same principle, we can always extend the quantum mechanical language (whatever it is supposed to be) to a language with the identity predicate for every quantum object. This is in fact true once we assume (i) that they are finite in number (which we may  in principle suppose), and (ii) they are  individuals, elements of sets. This last assumption is precisely what we are trying to suggest should avoided in order to keep (apparently) closer to quantum physics itself. But let us wait a moment, and go back to this point later. 

According to Saunders and Muller (Saunders [2006], Muller \& Saunders [2008]),  $a$ and $b$ are \textit{relational} if they satisfy  an irreflexive (for simplicity, binary) relation $R$. A `physically'  interesting example is in fact two electrons in a singlet state, which obey the irreflexive and symmetric relation ``$\ldots$ to have spin opposite from $\ldots$". The existence of this relation or of similar ones would be enough to force electrons, and fermions in general, to obey PII-R, and hence the criticism advanced by some authors (specifically French \& Krause, whom they mention) that they would violate PII would be misleading, according to Muller \& Saunders. But the argument above shows that Saunders' \textit{relational} objects are also individuals, and although he and Muller are right in saying ([2008], p.529) that ``absolute discernibles are always weak discernibles", they fail to recognize the absolute individuality of these entities, which is due to the underlying logic. As we have seen, the converse of their claim also holds, that is, relational objects are individuals---absolute discernibles. So, if fermions (in finite number) are to obey the classical theory of identity, they are able to bear proper names (according to the standard theory of direct reference), and to be ordered (in the finite case, this is trivial, but the general case needs  the axiom of choice or even something weaker), and so on. Are we really prepared for such consequences with regard to quantum entities, and for fermions in particular? I doubt that. Two fermions, electrons, say, in an anti-symmetric singlet state, yet obeying Pauli's principle---having opposite spins, cannot be individualized according to the theory of reference, for we never can say which is which. In particular, if they receive names (such as Mike and Ike, or Hans and Karl, to follow the example Muller \& Saunders take from Weyl), these names would be not proper names at all (rigid designators).  Thus, if we accept that the \emph{standard} concept of identity is what is lacking in sense for quantum entities, of course they do not obey PII, for it cannot even be formulated as usual. It is important not to say that it is not the case that these entities `violate'  PII; this is only a manner of speech, since the principle cannot even be formulated without identity.\footnote{An analogy perhaps is useful here. It is common to say that Birkhoff and von Neumann, in their celebrated paper from 1936, have proven that the distributive law is violated in QM. I think that perhaps the right way to speak of their result is that this law simply is not obeyed by the quantum connectives, for they are distinct from the classical ones. Thus, it is not the case that the distributive law is violated, but that it simply does not apply. The same happens in this case study.} I think that, just as the standard literature reminds us that we need to take care in transposing `microscopic' facts, such as the well known example given by Schr\"odinger's cat, to `macroscopic' cases, we should take the same care in transferring our usual speech in terms of `Hans' and `Karl' to electrons and the like. 

Very well, you may say that identity in quantum physics is not logical identity. But, then, what is it? You could say that the relation to be considered is not true identity, namely, the concept which would correspond to the diagonal of the domain, but a weaker relation, say a congruence relation of indiscernibility or Hilbert--Bernays identity. You might  also wish to avoid considering the `problematic' property $I_a$ above and consider just a weaker relation of indiscernibility. In this case, you are moving in the direction of accepting  quasi-set theory (French \& Krause [2006], chap.7).  There, the primitive relation is a congruence (not identity) $\equiv$ and we interpret $a \equiv b$ as ``$a$ is indistinguishable (or indiscernible) from $b$".  The theory admits the existence of two kinds of \textit{Urelemente}, which are termed $m$-objects and $M$-objects respectively. The last ones act as the \textit{Urelemente} of ZFU (the Zermelo--Fraenkel set theory with atoms), while the former are thought as representing non-individual quantum objects. 

There is also a defined concept of equality which has all the properties of classical identity when the involved objects are not $m$-objects, and the theory is articulated so that there is a copy of ZF (and of ZFU), where all standard mathematics can be erected. Thus the theory does not `destroy'  either classical logic or standard mathematics, but enlarges their scope, enabling us to talk of collections (quasi-sets) of really indiscernible objects, which (by the postulates of the theory) may have a cardinal number, but not an associated ordinal. In the `classical'  part of the theory, we can continue to do everything we do in standard ZF, including talking  about identity and diversity of certain objects (the `classical' ones). I shall not summarize the theory here, but just suggest  that it may be a more adequate language to speak of quantum objects; perhaps it is a candidate for the true language of quantum physics, even of QFT  (cf. French \& Krause op. cit., chap.9).

Of course we can also say that physical objects are not mere collections of entities, and I agree. The existence of isomers seems to prove that. Thus, a collection of certain objects must be characterized structurally, and I guess that, in what concerns mathematics, we can approach them by using the \textit{Urelemente} in quasi-set theory. Thus, a kind of mereology would be in order, a mereology where the parts may be indistinguishable and where the parts do not properly `sum'  to the wholes as in the usual (extensional) mereologies (Simons [1987]), for we have to respect quantum holism among other things. But this is a topic for further developments, for we need to overcome lots of difficulties indeed.  So, I guess that although identity is a quite useful concept, without which we probably would not be able to communicate our results and explain ourselves, it is not absolutely necessary at the level of the object language, and perhaps it is even problematic when used in the quantum realm. From the foundational point of view, if quantum objects (particles, fields, whatever you mean by quantum objects) are to be indistinguishable, they should be considered as so from the beginning, and not \textit{made} indiscernible by some mathematical device, such as the restriction of properties/relations  to what is inside the walls of some structure. What we really seem to use in our `quantum talk' is that we have certain quantities of elements of certain sorts, without taking them as individuals. The `cardinal talk' might be the solution, and when we say, for example, that `this'  electron here is distinct from `that' electron there, this might to be regarded as a manner of speech that has no counterpart in the formalism, as when we say that the structure $\mathcal{V}$ presented above is a `model' of ZF, even without being able to express it in terms of ZF proper.

\section*{Acknowledgments}
I would like to thank David Miller for comments and suggestions. The faults  that remain are of course mine.

\end{document}